\documentclass[runningheads]{llncs}
\usepackage{subcaption}
\usepackage{changepage}
\usepackage[T1]{fontenc}
\renewcommand{\subparagraph}{} 
\usepackage{titlesec}

\titlespacing*{\section}{0pt}{18pt}{6pt}
\titlespacing*{\subsection}{0pt}{14pt}{6pt}
\titlespacing*{\subsubsection}{0pt}{12pt}{6pt}

\usepackage{booktabs}
\usepackage{amsmath}
\usepackage{orcidlink}

\titlespacing*{\section}{0pt}{3.5ex plus 1ex minus .2ex}{0.5ex plus .1ex}

\usepackage{graphicx}
\begin{document}
\title{Exploring Runtime-Configurable Piecewise Linear Function Interpolation for FPGA-Based LLM Hardware Acceleration}
\titlerunning{Runtime-Configurable PWL in LLM FPGA-Based HW Acceleration}
%
\author{ 
Patrick Hugo Nepveu Nelson \orcidlink{0009-0007-5151-3286}\ \inst{1} \and
Názareth Fabiola Jiménez Chacón\orcidlink{0009-0005-3465-1738}\inst{1}\and
Roger de Jesús Morales Monge\orcidlink{0009-0009-2438-1745}\inst{1}\and
Luis~G.~Le\'on-Vega\orcidlink{0000-0002-3263-7853}\inst{1}\and
Jorge~Castro-God\'{i}nez\orcidlink{0000-0003-4808-4904}\inst{1}
}

\authorrunning{P. Nepveu et al.}

%
\institute{Costa Rica Institute of Technology, Cartago, Costa Rica
}
\maketitle              
\begin{abstract}
Non-linear functions constitute a significant computational cost in modern Transformer-based AI inference. While FPGAs improve efficiency through custom numerical representations, most implementations rely on function-specific hardware that limits flexibility and cross-model reuse. This work presents an FPGA accelerator featuring a specialized hardware datapath that evaluates multiple non-linear functions (exponential, GELU, ReLU, Sigmoid, Swish (SiLU), and tanh) via runtime coefficient configuration. Implemented on an AMD Alveo U55C FPGA, the architecture achieves fixed-point approximation errors on the order of $10^{-3}$ at 24-bit and 32-bit precisions. Integrating the framework into the \texttt{Qwen2.5-0.5B-Instruct} inference pipeline demonstrates that Q16.16, Q8.16, and Q8.8 configurations preserve stable language generation behavior, increasing cross-entropy loss by $\approx 0.083$ and perplexity from 11.31 to 12.29 relative to the FP32 baseline, while Q4.4 causes numerical failure. The results demonstrate that the proposed non-uniform piecewise-linear approach provides a practical balance between numerical fidelity, resource utilization, and architectural flexibility for reusable neural network acceleration. 
\end{abstract}

\section{Introduction}
Large Language Models (LLMs) based on Transformer architectures have become central to modern Artificial Intelligence workloads, enabling applications such as natural language processing, reasoning, and multimodal integration \cite{vaswani2017attention,haenlein2019brief}. Although Transformer inference is dominated by dense matrix operations, non-linear operators such as softmax and activation functions are executed repeatedly throughout every Transformer layer. While individually less computationally intensive than matrix multiplications, these operators rely on exponentiation, normalization, and other non-linear computations that are difficult to implement efficiently on FPGA architectures, often requiring dedicated hardware resources or iterative approximation methods \cite{wei2020softmax}.

Current inference platforms rely primarily on CPUs and GPUs. However, non-linear computations often exhibit poor hardware utilization because they involve data dependencies, iterative execution, and complex functions. As model sizes continue to increase, the cost of these operators becomes increasingly significant in both latency and energy consumption \cite{kachris2025survey}. Field-Programmable Gate Arrays (FPGAs) provide an attractive alternative for accelerating these workloads because they enable custom datapaths, arbitrary-precision arithmetic, and fine-grained parallelism. Other FPGA implementations of non-linear operators have explored polynomial approximations, lookup table (LUT) interpolation, and CORDIC-based methods~\cite{leiva-softmax,carrillo2025cordic_softmax}. Nevertheless, many existing designs remain tightly specialized to a single function, limiting flexibility across different inference workloads. This work presents a configurable FPGA acceleration framework for Transformer non-linear operators based on piecewise-linear interpolation. The proposed architecture relies on a Python-based coefficient-generation tool that applies non-uniform segmentation to a target function and automatically produces the interpolation data. This allows the same hardware to support different non-linear functions without requiring architectural modifications. The accelerator is integrated into a heterogeneous CPU-FPGA runtime using PyXRT and dynamic operator interception, enabling transparent deployment within PyTorch Transformer pipelines without modifying the original model implementation \cite{amd_xilinx_xrt}.

To evaluate the proposed framework, both the Swish activation function and softmax attention operators are intercepted during runtime and redirected to an AMD Alveo U55C platform. The evaluation combines hardware-level latency characterization with end-to-end language modeling analysis using WikiText-2 and autoregressive text generation workloads. The main contributions of this work are: (1) a runtime-configurable FPGA accelerator for non-linear operators using piecewise-linear interpolation and parameterized fixed-point arithmetic,
    (2) a heterogeneous runtime integration framework based on PyXRT and dynamic operator interception for transparent deployment within PyTorch inference pipelines,
    (3) a comprehensive numerical fidelity evaluation using attention-divergence metrics, perplexity analysis, and autoregressive text generation workloads, and (4) a detailed system-level latency analysis, identifying PCIe communication and host-device synchronization as dominant bottlenecks in heterogeneous FPGA inference.

The remainder of this paper is structured as follows: Section 2 presents the theoretical background and related work on non-linear function approximation methods, fixed-point arithmetic and FPGA-based acceleration. Section 3 describes the proposed FPGA accelerator architecture, the piecewise-linear approximation framework and the heterogeneous runtime integration using PyXRT. Section 4 presents the experimental methodology and discusses the hardware and application-level evaluation results. Finally, Section 5 summarizes the main findings of this work and discusses future research directions.

\section{Background and Related Work}
FPGAs provide a flexible platform for accelerating Transformer inference because they enable custom datapaths, pipeline parallelism, and arbitrary-precision arithmetic \cite{kastner2018parallel}. Unlike CPUs and GPUs, FPGA architectures allow hardware resources to be tailored to the numerical requirements of a specific workload, making them particularly attractive for non-linear function evaluation under reduced-precision arithmetic~\cite{llm_fpga}. In practice, fixed-point representations are commonly used to reduce hardware complexity, latency, and Digital Signal Processing (DSP) cell utilization. However, decreasing numerical precision introduces quantization noise and clipping effects that can propagate through Transformer inference pipelines and alter model behavior \cite{gupta2015deep}.

Additionally, the use of approximate computing reduces hardware accelerator resource consumption and latency, thanks to the probabilistic nature of AI models such as deep neural networks and LLMs~\cite{leiva-softmax}. This has led to multiple proposals for implementing non-linear functions on FPGA platforms. One common approach uses domain-limited Taylor series expansions, in which transcendental functions are approximated using low-order polynomial arithmetic. These methods provide low-latency implementations because they primarily rely on additions and multiplications, which map efficiently onto FPGA DSP resources. However, approximation accuracy degrades rapidly outside the function's domain, often requiring aggressive clipping or range restriction \cite{leiva-softmax,muller2016elementary}. LUTs represent another widely used approach. In these implementations, the non-linear function is sampled offline and intermediate values are reconstructed via piecewise-linear interpolation. Linear and quadratic interpolation methods typically achieve higher numerical accuracy than low-order polynomial approximations at the expense of increased memory utilization and additional interpolation logic \cite{lachowicz2008fpga}. Piecewise linear interpolation is particularly attractive for FPGA architectures because it provides a favorable trade-off between arithmetic complexity, approximation accuracy, and hardware efficiency \cite{wei2020softmax}.

CORDIC-based implementations provide an alternative approach by replacing multipliers with iterative shift-add operations. These methods significantly reduce DSP utilization and are, therefore, attractive for resource-constrained FPGA deployments \cite{carrillo2025cordic_softmax}. Nevertheless, their iterative execution behavior generally introduces higher latency compared to LUT-based interpolation schemes \cite{mahbub2026hardware_journal,carrillo2025cordic_softmax}.

On the other hand, modern and High-Performance Computing (HPC)-oriented FPGA development increasingly relies on High-Level Synthesis (HLS), which enables hardware generation from untimed high-level programming languages such as C/C++ \cite{xilinx_hls}. HLS simplifies accelerator development while still allowing explicit control over hardware optimizations including pipelining, loop unrolling, and memory partitioning \cite{amd_dataflow_canonical}, while improving development productivity through higher-level language abstractions. These optimizations are particularly important for non-linear function acceleration because throughput is often constrained by memory bandwidth, arithmetic dependencies and irregular dataflow patterns \cite{kachris2025survey}. Besides, FPGA architectures enable multiple forms of parallelism, including spatial parallelism, pipeline parallelism, and data-level parallelism \cite{kastner2018parallel}. By exploiting these characteristics, non-linear operators can be implemented in deeply pipelined streaming architectures that process new inputs every clock cycle. This flexibility makes FPGA platforms well-suited for customized Transformer inference accelerators where numerical precision, memory organization, and datapath structure must be jointly optimized \cite{chen2025fpga}.

Overall, research on reconfigurable non-linear hardware accelerators has been limited. Most solutions focus on function-specific hardware accelerators that are only configurable at synthesis time, limiting the use of these accelerators for other functions that may appear across multiple models. This creates an opportunity to study runtime-configurable accelerators that can be reused across multiple functions and models, allowing the accelerator to compute multiple functions by simply loading coefficients into LUT-style memory, together with tools like HLS to enable AI-acceleration faster than traditional FPGA development flows.

\section{Configurable FPGA Accelerator Architecture}
This work proposes an implementation of a runtime function-configurable FPGA accelerator described in HLS. Our design follows a traditional streaming dataflow model composed of three stages: input loading, computation, and output storage. These stages operate concurrently using the \texttt{HLS DATAFLOW} directive, enabling overlap between communication and computation. To sustain continuous streaming execution, loop pipelining was applied throughout the load, compute, and store stages using the \texttt{HLS PIPELINE II=1} directive. At the core of the architecture lies a shared PWL compute engine composed of replicated interpolation lanes operating in parallel under a streaming dataflow execution model. Each lane evaluates the piecewise-linear approximation 
\begin{equation}
f(x) \approx m_i(x-x_{0,i})+b_i
\end{equation}
where $m_i$, $b_i$ and $x_{0,i}$ correspond to the slope, intercept, and breakpoints associated with segment $i$.

Our architecture supports two types of configuration for its approximation parameters. Numerical precision and vectorization widths are synthesis-time configurable via a C++ header file before bitstream generation. Meanwhile, the operating range and execution mode are runtime-configurable via AXI-Lite transactions between the CPU and the FPGA. Prior to kernel execution, the host transfers the segment mapping, breakpoints, slopes, and intercepts associated with the selected function into persistent on-chip lookup memories. Once loaded, these coefficients are reused across subsequent kernel invocations, allowing the accelerator to evaluate different non-linear functions without requiring hardware resynthesis. Changing the approximated function therefore only requires updating the coefficient tables rather than modifying the hardware implementation.

When designing a LUT-based accelerator, a bottleneck arises from concurrent access to the segment mapping to memory. When multiple elements of the vector are resolved simultaneously, the memory experiences multiple concurrent accesses, leading to multiple consumers accessing the same memory. In particular, using a single shared LUT introduces memory conflicts that prevent II=1 execution. Our implementation eliminates this bottleneck by replicating lookup structures across compute lanes via array partitioning and memory banking, using the \texttt{HLS ARRAY\_PARTITION} directive, allowing every interpolation lane to access the required coefficients independently.

\subsection{Piecewise Linear Approximation Engine}
The core of the accelerator is a non-uniform piecewise-linear approximation engine designed to efficiently approximate non-linear functions in fixed-point arithmetic. The target function's domain is divided into non-uniform segments following a greedy adaptive strategy that iteratively selects the largest interpolation interval that satisfies a predefined approximation error constraint. This improves approximation accuracy in regions with high curvature while minimizing the total number of segments. The segmentation strategy is derived from the Lagrange remainder error formulation for linear interpolation:

\begin{equation}
    R(x)=\frac{f''(\xi)}{2}(x-x_i)(x-x_{i+1})
\end{equation}

\noindent where $R(x)$ represents the interpolation error at point $x$, $f''(\xi)$ is the second derivative of the target function evaluated at some intermediate point $\xi \in [x_i,x_{i+1}]$, and $x_i$, $x_{i+1}$ denote the boundaries of the interpolation interval \cite{burden2015numerical}. 

Directly searching among irregular interval boundaries in hardware would require multiple comparison operations, increasing latency and logic utilization. To eliminate this overhead, the proposed architecture converts the irregular segmentation into a dense, uniform mapping table during runtime configuration preprocessing using a dedicated Python-based framework. To maximize hardware efficiency, two key hardware-oriented optimization constraints were used during this discretization process. First, the total operating range is discretized using a power-of-two number of total bins. Second, individual segment widths are strictly constrained to powers of two. By enforcing these dual power-of-two constraints, the complex control logic traditionally required for segment indexing is removed. Thus, instead of relying on costly hardware division or iterative boundary searches, the segment index calculation reduces to a trivial bit-shifting operation.

\subsection{Unified Compute Engine for Point-wise and Softmax Execution}
\begin{figure}[h!]
    \centering
    \includegraphics[width=1\linewidth]{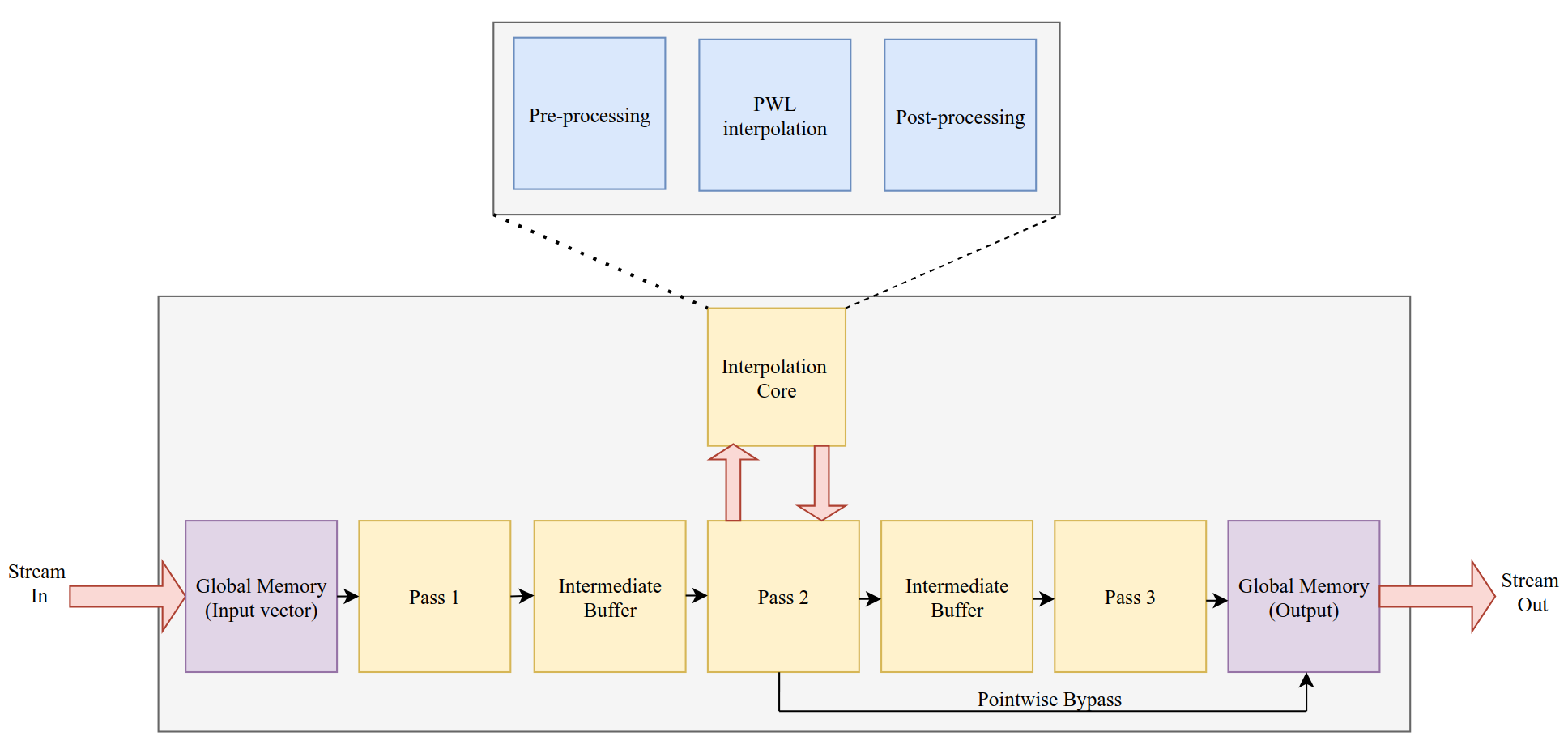}
    \caption{Proposed Accelerator Architecture. The three passes are pipelined and part of the compute stage. Additionally, the interpolation core is pipelined to follow a dataflow execution pattern.}
    \label{fig:system}
\end{figure}

Fig.~\ref{fig:system} depicts the overall architecture of the proposed accelerator. The design employs a unified compute engine that supports both point-wise non-linear evaluation and softmax computation on the same underlying PWL interpolation hardware. The interpolation core consists of three stages: pre-processing, PWL interpolation, and post-processing, while the surrounding datapath orchestrates the execution flow according to the selected operating mode. This shared-resource architecture represents one key design decision, as it avoids duplicating non-linear compute units while maximizing hardware utilization. The compute engine operates under a mode-dependent execution scheme controlled at runtime via AXI-Lite control registers, exposed via the \texttt{HLS INTERFACE s\_axilite} directive. This enables the CPU to dynamically configure the accelerator at runtime by selecting the execution mode and loading the interpolation tables corresponding to the selected non-linear function without requiring hardware resynthesis. The interpolation tables are loaded once before execution and reused across subsequent kernel invocations until a new function is selected. In point-wise mode, the datapath directly evaluates the selected non-linear function using the runtime-loaded interpolation tables stored in on-chip memory. For softmax, the architecture reuses the same interpolation hardware to implement the SafeSoftmax formulation, a numerically stable variant widely adopted in modern Transformer accelerators and deep learning frameworks \cite{mahbub2026hardware_journal}. Because SafeSoftmax introduces inter-element dependencies and normalization operations, the accelerator executes softmax using a three-pass pipeline.

As illustrated in Fig.~\ref{fig:system}, the first pass computes the global maximum of the input vector through a parallel reduction tree across multiple lanes, lowering the time complexity from $O(n)$ to $O(\log_2 n)$. During the second pass, the shared PWL datapath is reused to evaluate the exponential approximation of each normalized input value, using the maximum value identified during the first pass to compute  $e^{x_i-\max(\mathbf{x})}$, while simultaneously accumulating the partial sums required for normalization. Similar to the maximum computation, the accumulation stage is implemented using a fully unrolled sum reduction tree, allowing additions to be performed in parallel and minimizing the reduction's latency. Finally, in the third pass, the accumulated exponential values are normalized by multiplying each element by the inverse of the global sum, producing the final softmax probability distribution. Before FPGA execution, tensors undergo quantization, packetization, and alignment padding in a pre-processing stage. This ensures compatibility with the 512-bit AXI memory interface while preserving deterministic DMA behavior, where rather than processing scalar values individually, each AXI transaction contains multiple packed fixed-point elements processed simultaneously across replicated compute lanes. The number of parallel lanes is determined by $k_{\text{Packets}}=\frac{{\text{BusWidth}}}{k{\text{PaddedWidth}}}$. 

\section{Results and Discussion}
The proposed accelerator was synthesized using AMD Vivado/Vitis 2024.1 targeting the AMD Alveo U55C FPGA platform with a target operating frequency of 200~MHz. The experiments were conducted on the HACC high-performance computing cluster at ETH Zürich. The host system consisted of a Dell PowerEdge R740 server running Ubuntu 22.04.5 LTS, equipped with dual Intel Xeon Gold 6248 processors (40 physical cores, 80 hardware threads total) operating at 2.50~GHz and 376~GB of system memory. The host-side software stack was implemented in Python 3.10.12 using PyTorch and the PyXRT bindings distributed with Xilinx Runtime (XRT) version 2.17.319 (Vitis 2024.1). A key objective of the runtime framework was preserving compatibility with existing Transformer software stacks without modifying the original model implementations. To achieve this, the runtime employs dynamic interception mechanisms using Python monkey-patching and custom context managers for activation-function hijacking and softmax interception. To evaluate the numerical fidelity of the FPGA accelerator, three complementary metrics are employed: Root Mean Square Error (RMSE), Kullback–Leibler (KL) divergence, and Signal-to-Quantization Noise Ratio (SQNR). These metrics are used to compare the outputs of the FPGA implementation with the FP32 software baseline and to quantify the numerical deviations introduced by fixed-point arithmetic and PWL approximation. At the application level, the evaluation additionally considers the effect of the accelerator on language-modeling behavior through Cross-Entropy Loss, Perplexity, Top-1 Token Alignment and Autoregressive Token Match.

\subsection{Hardware Performance Benchmark}
To evaluate the hardware efficiency of the proposed architecture, it was compared against three FPGA-oriented implementations: CORDIC-based exponentiation, uniform LUT interpolation, and third-order Taylor approximation \cite{leiva-softmax,carrillo2025cordic_softmax}. The CORDIC approach computes transcendental functions through iterative shift-and-add rotations. The uniform LUT method stores precomputed function samples and performs interpolation over equally spaced intervals. While computationally simple, its accuracy depends directly on the number of stored entries, causing memory requirements to increase rapidly as precision requirements become more stringent. The third-order Taylor approximation evaluates a polynomial expansion of the target function, reducing memory usage but increasing DSP consumption due to the multiple multiplication stages required for polynomial evaluation.

\begin{figure}[!h]
    \vspace*{-1cm}
    \centering
    \includegraphics[width=1\linewidth]{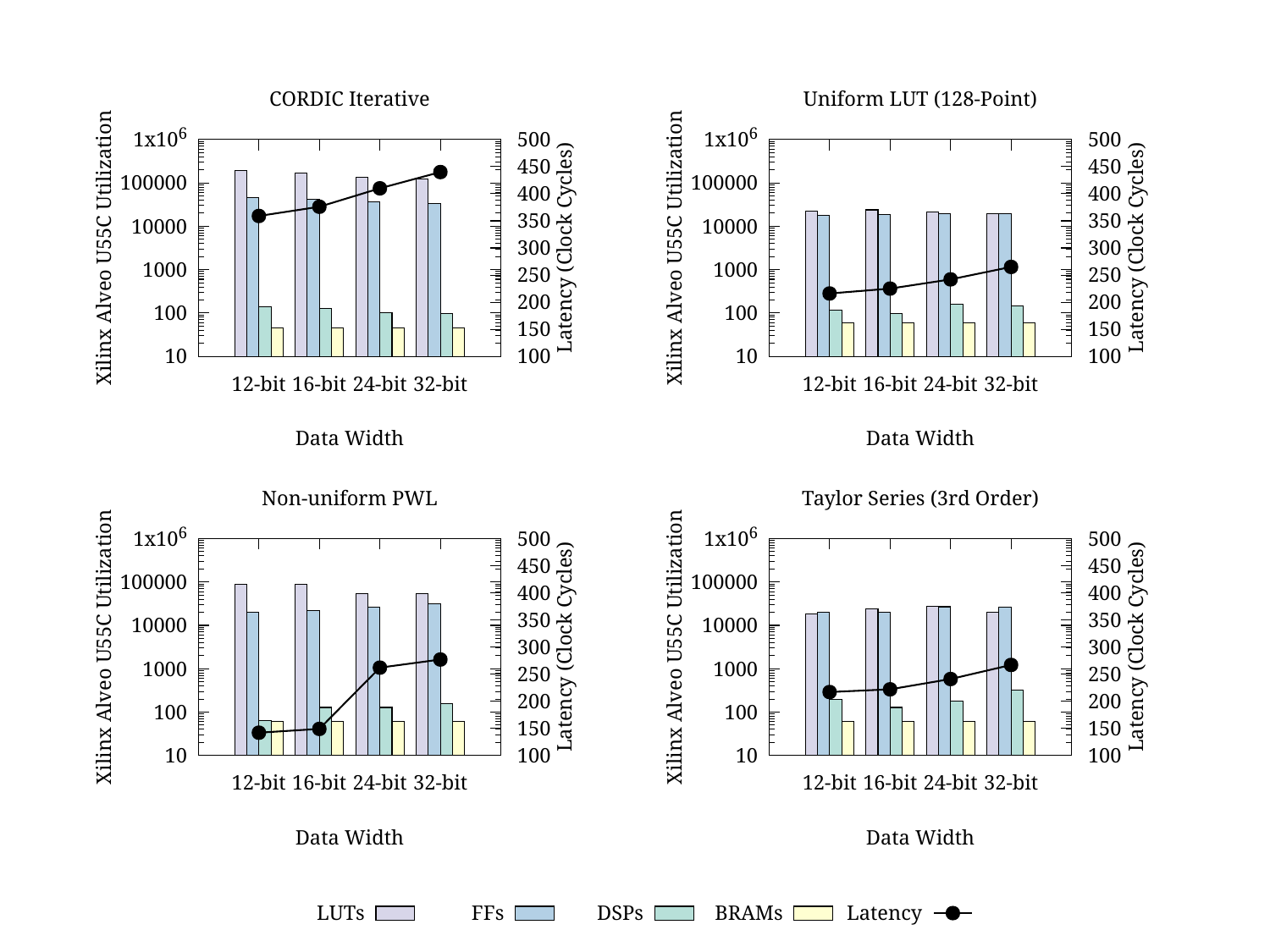}
    \caption{Resource utilization and execution latency of the evaluated softmax accelerators across fixed-point precisions from 12 to 32 bits for a vector size of 1024 elements operating at 200~MHz. The proposed architecture corresponds to the non-uniform PWL approximation engine.}
    \label{fig:resources}
\end{figure}

Fig.~\ref{fig:resources} presents the resource utilization and execution latency obtained for fixed-point precisions ranging from 12 to 32 bits. At 16-bit precision, the proposed accelerator achieves the lowest execution latency among all evaluated architectures, requiring only 149 clock cycles. In comparison, the CORDIC implementation requires 376 cycles, while the uniform LUT and third-order Taylor approaches require 225 and 222 cycles, respectively. This corresponds to a latency reduction of approximately 60.4\% relative to CORDIC and roughly 34\% relative to both the LUT-based and Taylor-based implementations. To ensure a fair evaluation of numerical fidelity under the same 16-bit precision, the softmax operation was profiled across an input range of $[-4,4]$. The third-order Taylor series achieved a global Softmax RMSE of $0.001954$, while the 128-point uniform LUT resulted in an RMSE of $0.001963$. In comparison, the iterative CORDIC approach yielded an RMSE of $0.001194$. The proposed non-uniform PWL engine demonstrated highly competitive accuracy, achieving an RMSE of $0.001215$. The profiling revealed that Taylor and uniform LUT configurations suffered from severe underflow, registering $0.0000$ accumulation outputs. From a resource-utilization perspective, the proposed architecture occupies 89,783 LUTs, 22,139 FFs, 128 DSPs and 61 BRAMs at 16-bit precision. Compared with CORDIC, LUT utilization is reduced by approximately 47\% while maintaining the same DSP usage. Although the uniform LUT and Taylor implementations require fewer LUT resources, they do not provide the flexibility of supporting arbitrary non-linear functions through runtime coefficient loading.

\subsection{Approximation Accuracy Across Non-linear Functions}
\begin{figure}
    \vspace*{-2.5cm}
    \centering
    \includegraphics[width=1\linewidth]{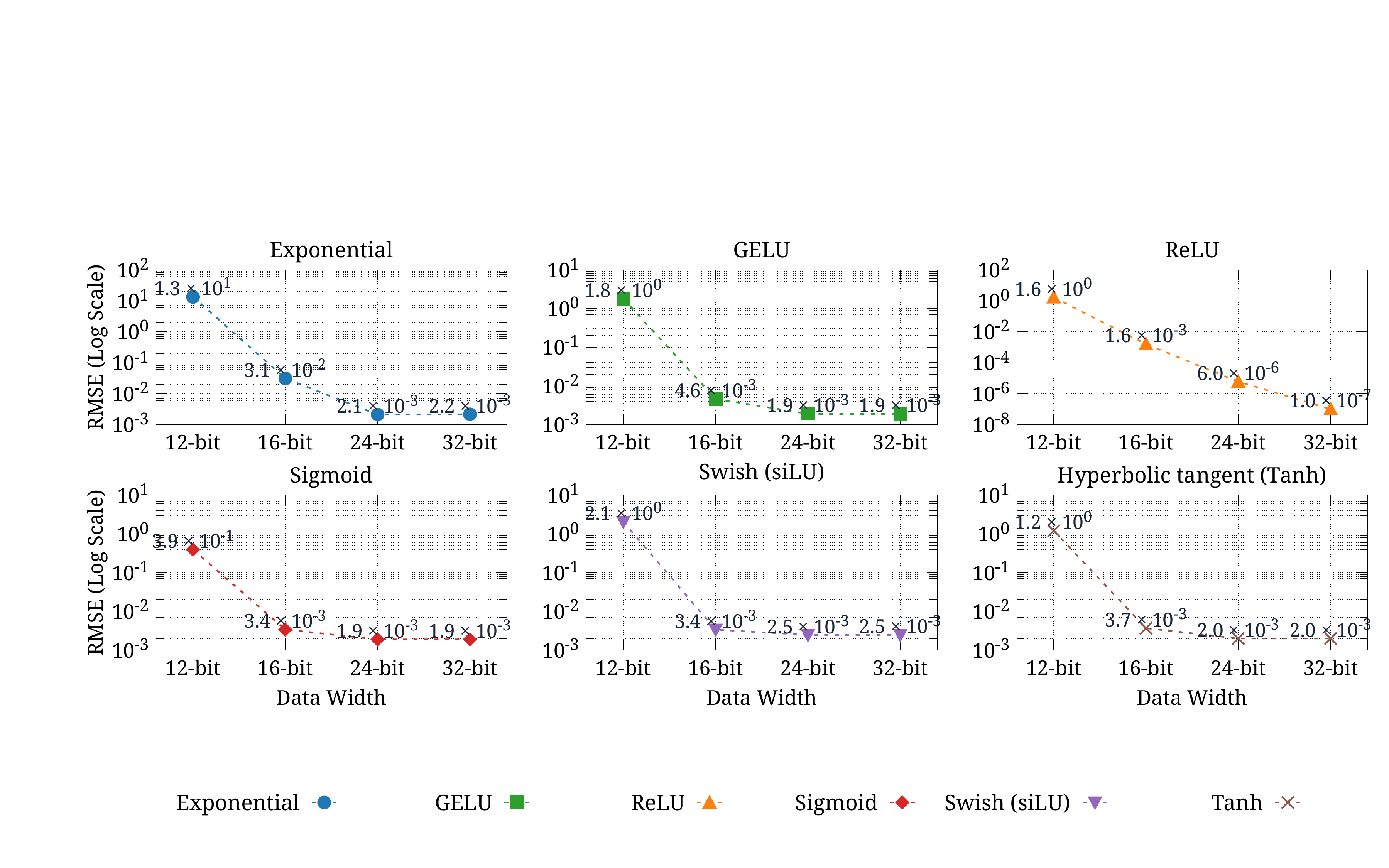}
    \caption{RMSE comparison of the multiple non-linear functions across multiple fixed-point precisions using input values within the range $[-4,4]$.}
    \label{fig:rmse_functions}
\end{figure}

To evaluate the generality of the proposed accelerator, the non-uniform PWL approximation engine was tested using several non-linear functions commonly employed in machine learning, including exponential, GELU, ReLU, Sigmoid, Swish, and tanh. Fig.~\ref{fig:rmse_functions} presents the RMSE obtained for different fixed-point configurations. To enable a fair comparison across all evaluated functions, the same input interval of  $[-4,4]$ was adopted. Furthermore, all evaluated formats employed an 8-bit integer partition, with the remaining bits allocated to the fractional component for consistency. The exponential function exhibited the highest sensitivity to quantization effects. At 12-bit precision, the RMSE reached approximately 13.3 due to the rapid growth of the function and the limited fractional resolution available. Increasing the precision to 16 bits reduced the RMSE by more than two orders of magnitude to $3.1\times10^{-2}$, while the 24-bit and 32-bit configurations converged near $2\times10^{-3}$. This behavior indicates that quantization dominates the error at low precision, whereas systematic error becomes the limiting factor at higher precisions.

The GELU, Sigmoid, Swish, and tanh functions showed similar trends, with RMSE values decreasing substantially between the 12-bit and 16-bit configurations and stabilizing around $10^{-3}$ for higher precisions. Their smoother behavior and bounded operating regions improve numerical robustness compared to the exponential function. The results demonstrate that the proposed architecture maintains consistent approximation accuracy across a diverse set of non-linear functions without requiring hardware modifications. Furthermore, the dominant error source transitions from fixed-point quantization at low precisions to interpolation fidelity at higher precisions, confirming the effectiveness of the non-uniform segmentation strategy for general-purpose non-linear function evaluation.

\subsection{Numerical Accuracy and Transformer Evaluation}
\begin{figure}[htbp]
    \vspace*{-2cm}
    \centering
    \begin{adjustwidth}{0cm}{0cm} 
        \begin{subfigure}[b]{0.50\textwidth} 
            \includegraphics[width=\linewidth]{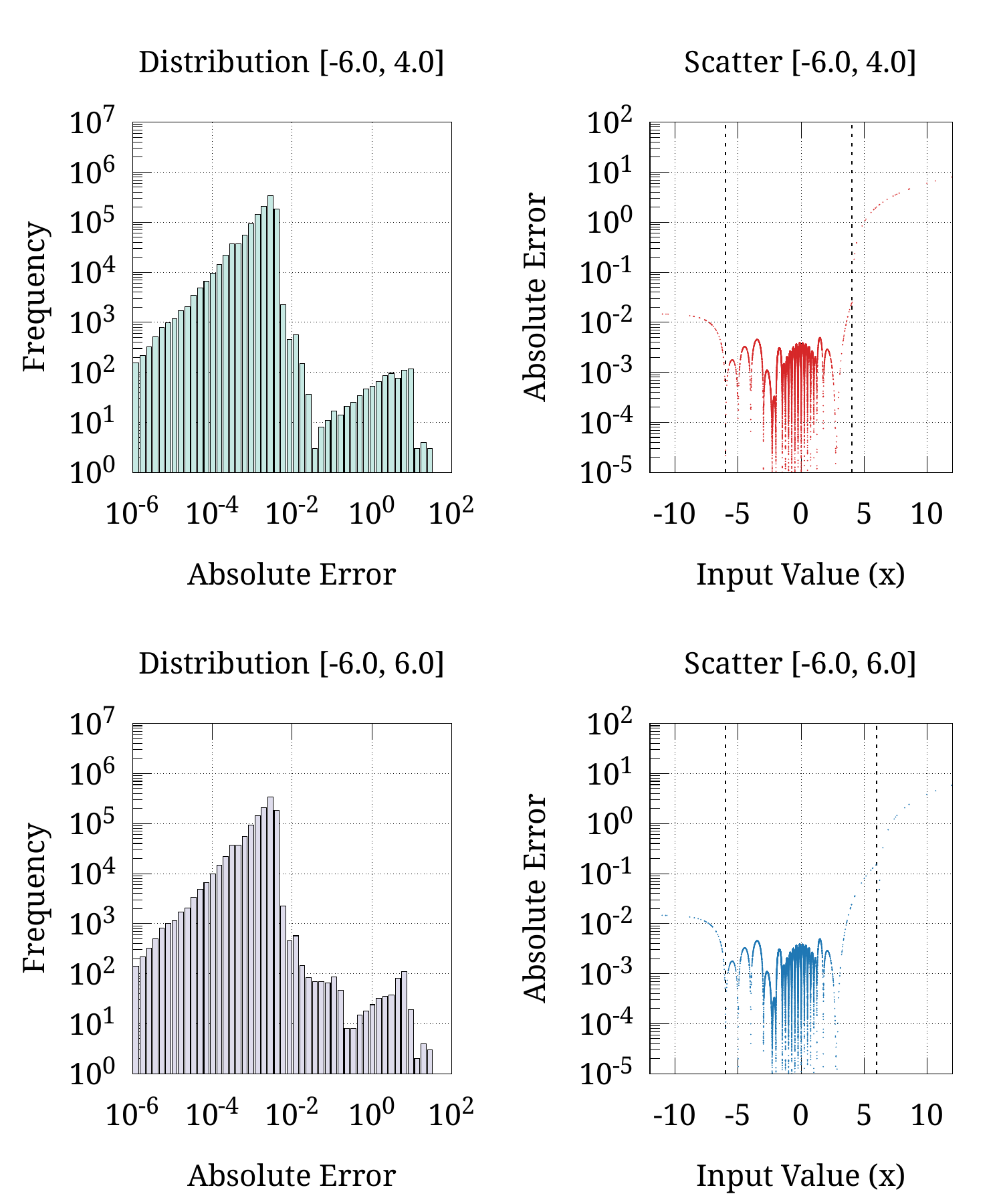} 
            \caption{Q16.16} 
        \end{subfigure}
        \begin{subfigure}[b]{0.50\textwidth} 
            \includegraphics[width=\linewidth]{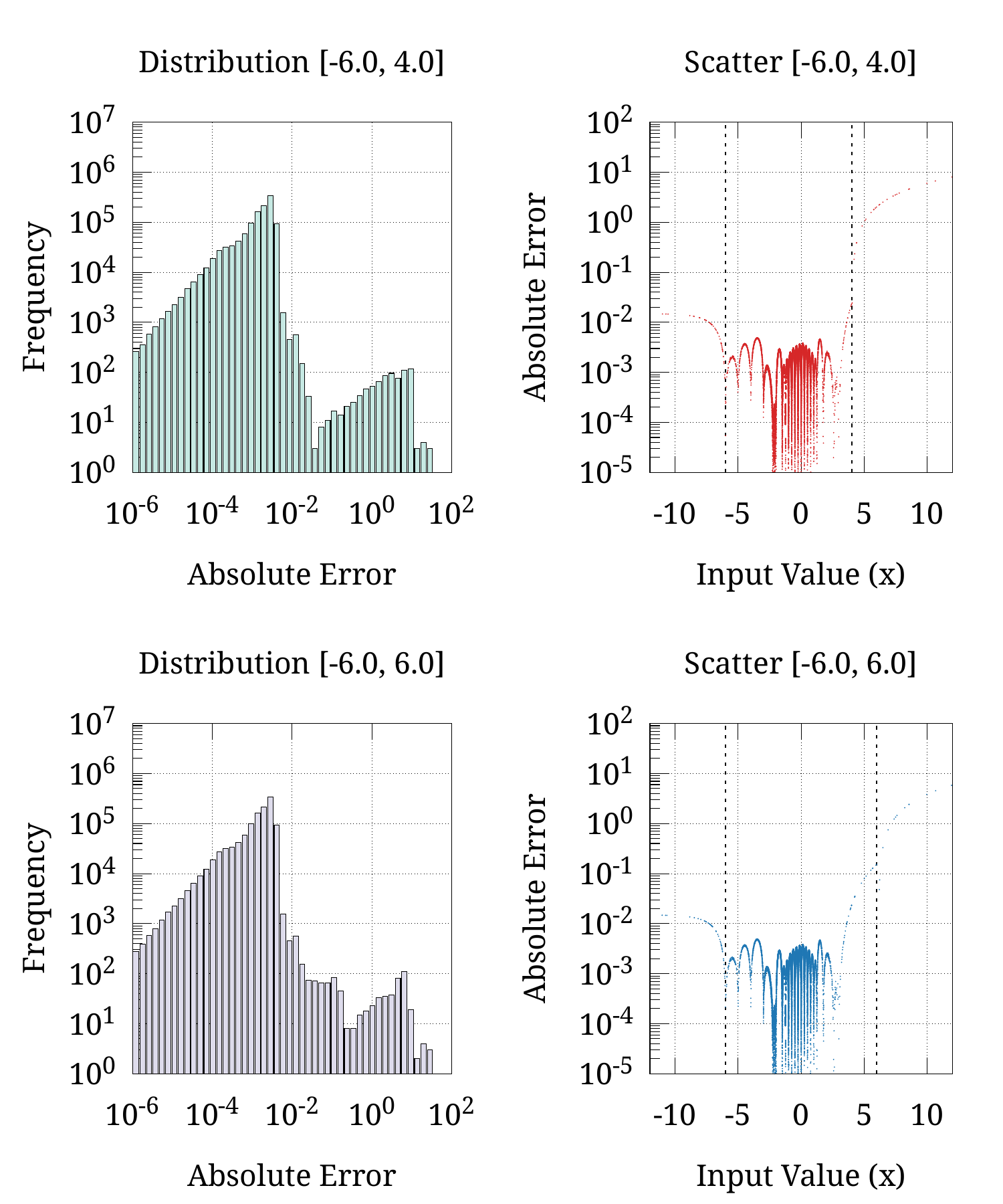} 
            \caption{Q8.16} 
        \end{subfigure}
        
        \vspace{1ex}
        
        \begin{subfigure}[b]{0.50\textwidth} 
            \includegraphics[width=\linewidth]{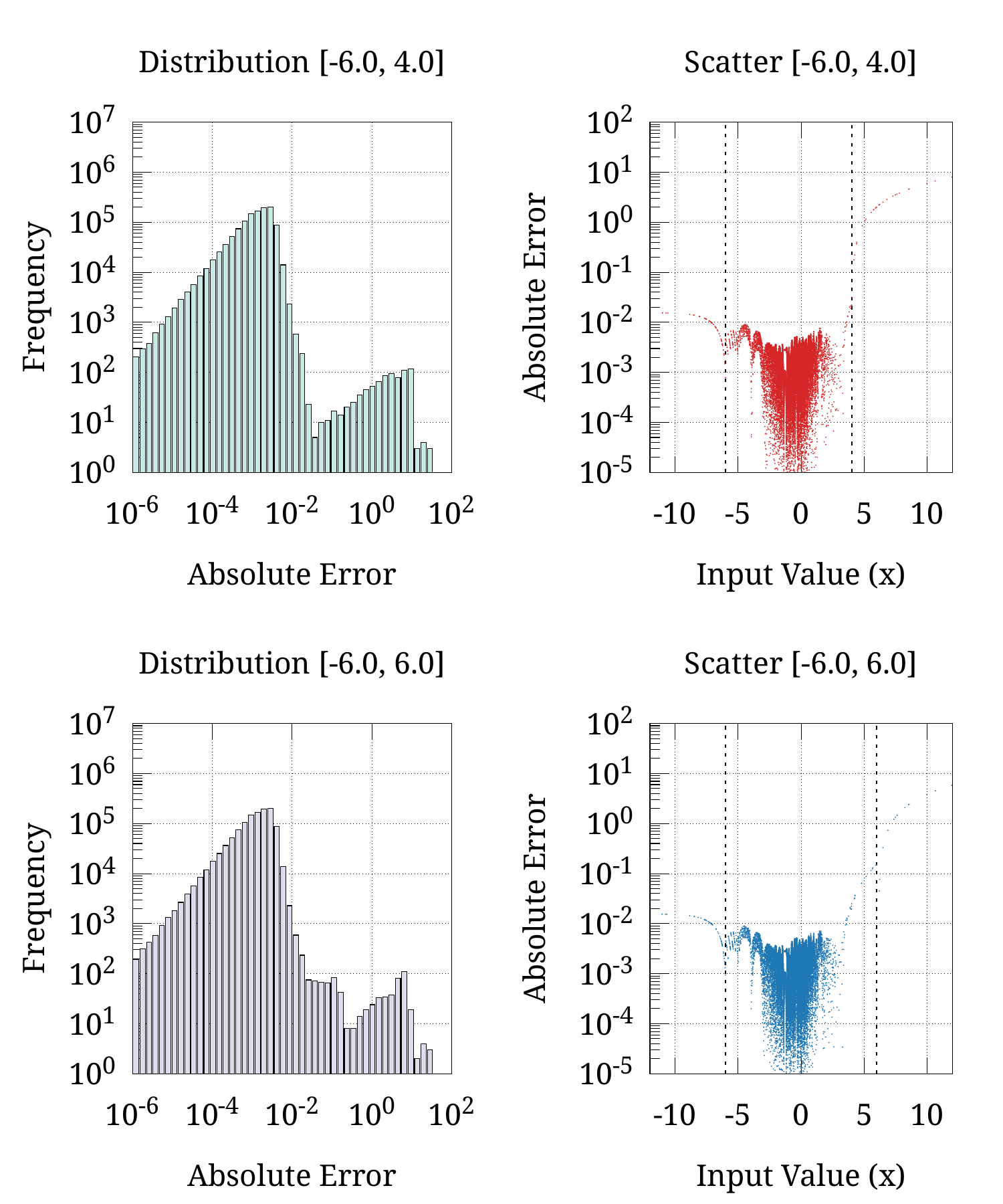} 
            \caption{Q8.8} 
        \end{subfigure} \hfill
        \begin{subfigure}[b]{0.50\textwidth} 
            \includegraphics[width=\linewidth]{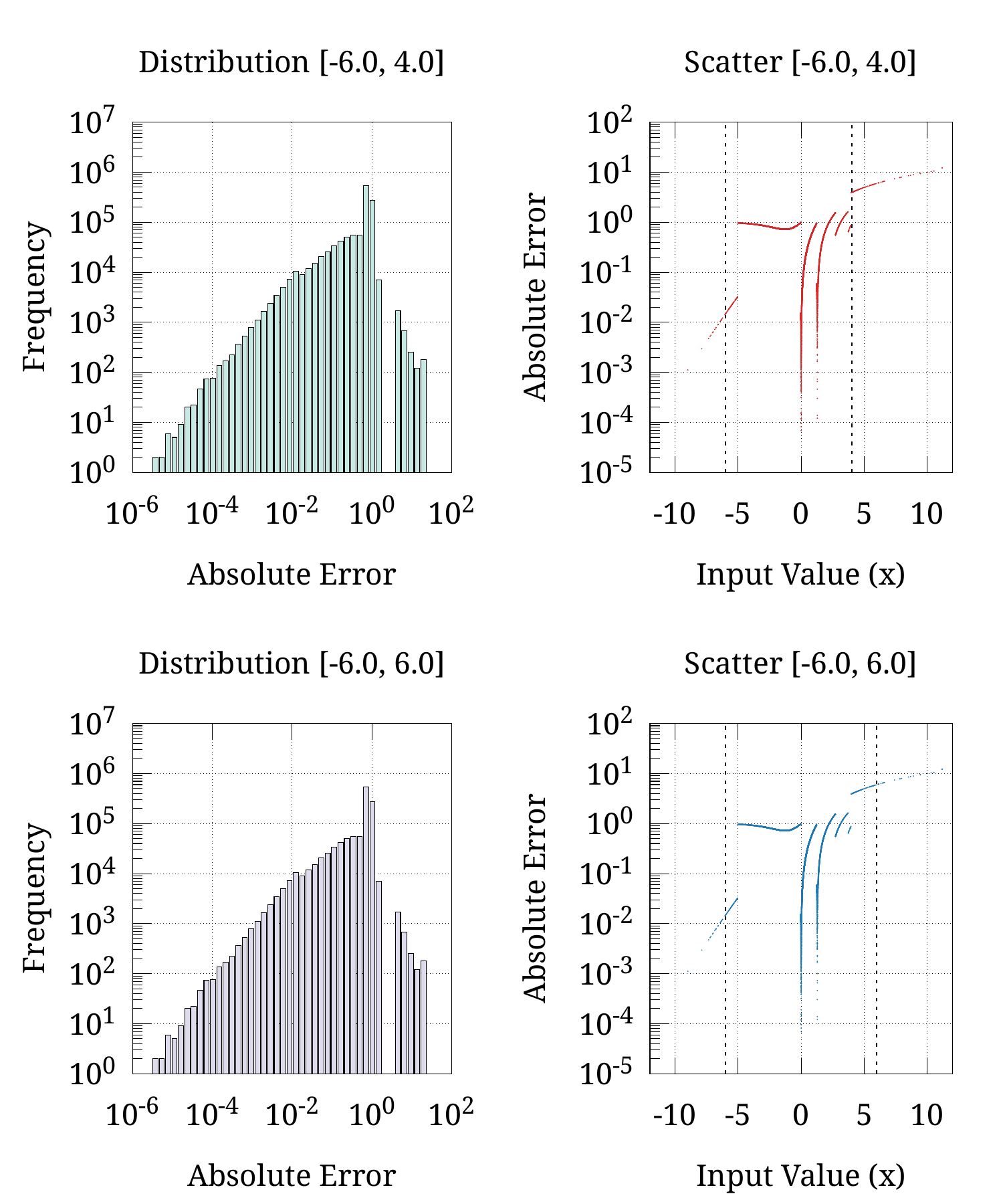} 
            \caption{Q4.4} 
        \end{subfigure}
    \end{adjustwidth} 

    \caption{Residual distributions obtained for the fixed-point configurations under both the standard interpolation interval $[-6,4]$ and the extended interval $[-6,6]$. Similar distributions suggest clipping of activation outliers rather than quantization noise.}
    \label{fig:histo} 
\end{figure}

To evaluate the numerical behavior of the proposed accelerator under realistic inference conditions, a layer-wise residual analysis was performed on the 24 Transformer MLP layers of the \texttt{Qwen2.5-0.5B-Instruct} model using the experimental framework described earlier. The analysis focused on the Swish (SiLU) activation path within the MLP submodules, where the proposed FPGA accelerator replaced the native FP32 implementation. Fig.~\ref{fig:histo} presents the residual distributions obtained for the Q16.16, Q8.16, Q8.8, and Q4.4 fixed-point configurations under both the standard interpolation interval $[-6,4]$ and the extended interval $[-6,6]$. The Q16.16, Q8.16, and Q8.8 configurations exhibit similar residual distributions despite their different fractional precisions. For these formats, most approximation errors remain concentrated below $10^{-2}$, indicating that the non-uniform PWL approximation preserves the overall shape of the Swish activation with limited numerical distortion. The similarity of the residual histograms suggests that clipping of activation outliers, rather than quantization noise, becomes the dominant source of error for these higher-precision configurations.

In contrast, the Q4.4 configuration exhibits a substantial increase in residual magnitude. The reduced fractional resolution introduces significant quantization effects, causing the error distribution to shift toward larger values and resulting in a noticeable degradation of approximation fidelity. This behavior indicates that the 8-bit representation is insufficient to accurately represent the interpolation coefficients and intermediate activation values required by the Swish function. The scatter plots further reveal that the largest residuals occur outside the interpolation domain, where activation values are clipped to the interval boundaries. Several Transformer layers generate activation outliers that exceed the supported interpolation range, producing a characteristic increase in approximation error. Because all evaluated formats share the same interpolation limits, this clipping behavior affects Q16.16, Q8.16, and Q8.8 similarly and explains the small RMSE differences observed between these configurations.

To investigate the impact of clipping, the interpolation interval was expanded from $[-6,4]$ to $[-6,6]$. The larger dynamic range reduces the number of clipped activations and improves approximation accuracy across multiple layers. The global RMSE decreases from 0.1366 to 0.1017, corresponding to a reduction of approximately 25.5\%. Similar improvements are observed in the layer-wise SQNR and maximum error metrics. These results indicate that the proposed accelerator maintains stable numerical behavior for Q16.16, Q8.16, and Q8.8 representations, while Q4.4 introduces excessive quantization error that significantly degrades approximation quality. Furthermore, the analysis shows that clipping of activation outliers represents the primary source of residual error at practical precision levels, suggesting that interpolation range selection has a larger impact on accuracy than further increases in arithmetic precision.

\subsection{Autoregressive Text Generation}
The practical impact of the proposed accelerator was evaluated using the same \texttt{Qwen2.5-0.5B-Instruct} Transformer model on the WikiText-2 benchmark. The analysis focused on the Transformer MLP submodules, where the Swish activation function was replaced by the FPGA-based non-uniform PWL approximation through the runtime framework. The FP32 software execution served as the reference baseline. First, a static sequence evaluation was conducted using the WikiText-2 benchmark, where the same input sequence was evaluated using both the original FP32 software baseline and the FPGA-accelerated activation function. Cross-Entropy Loss and Perplexity were computed directly from the model outputs. To quantify distributional deviations introduced by the approximation, the KL divergence between the vocabulary probability distributions of both executions was measured. Additionally, Top-1 Token Alignment was computed as the percentage of positions for which both implementations selected the same most-probable token. Second, autoregressive text generation was evaluated using deterministic greedy decoding. The model generated 100 new tokens from a fixed prompt, where token sequences were compared position by position to compute the Autoregressive Token Match metric, which measures the percentage of generated tokens that remain identical throughout the decoding process. Table~\ref{tab:wikitext_results} summarizes the quantitative evaluation metrics. The Q16.16, Q8.16, and Q8.8 configurations exhibit nearly identical behavior, with cross-entropy degradation limited to approximately $\Delta L \approx 0.083$, perplexity increasing by less than one point relative to the FP32 baseline and Top-1 Token Alignment remaining close to 85\%. Similarly, the KL divergence remains low ($\approx0.093$), indicating that the approximated activations preserve the output distribution of the original model.

\begin{table}[!t]
\vspace*{-0.5cm}
\centering
\caption{Quantitative evaluation on the WikiText-2 benchmark.}
\label{tab:wikitext_results}
\begin{tabular}{lccccc}
\toprule
Configuration & Loss ($L$) & $\Delta L$ & Perplexity & Top-1 Align. (\%) & KL Div. \\
\midrule
SW Baseline (FP32) & 2.4257 & -- & 11.3104 & 100.00 & 0.0000 \\
FPGA Q16.16 & 2.5092 & +0.0835 & 12.2948 & 85.10 & 0.0932 \\
FPGA Q8.16 & 2.5090 & +0.0833 & 12.2924 & 85.14 & 0.0933 \\
FPGA Q8.8 & 2.5090 & +0.0832 & 12.2922 & 85.10 & 0.0937 \\
FPGA Q4.4 & 16.5250 & +14.0993 & 15,021,856.29 & 0.00 & 14.3490 \\
\bottomrule
\end{tabular}
\end{table}

In contrast, the Q4.4 configuration exhibits severe numerical degradation. The loss increases to 16.53, perplexity exceeds $1.5\times10^{7}$, and Top-1 Token Alignment collapses to 0\%. This behavior is attributed to the limited fractional precision and dynamic range of the 8-bit representation, which introduce substantial quantization and clipping errors within the Swish approximation. To evaluate recursive error propagation, autoregressive generation was performed using deterministic greedy decoding over a 100-token window. Although the token-level match rate decreases to approximately 9\% for Q16.16 and Q8.16, and 7\% for Q8.8, the generated sequences remain grammatically coherent and semantically consistent with the FP32 baseline. The observed divergence originates from small perturbations in early decoding steps that alter the subsequent context history and progressively shift the decoding trajectory. The Q4.4 configuration fails to maintain coherent generation and produces repetitive or corrupted token sequences, indicating a complete breakdown of the Transformer computation pipeline. The results demonstrate that Q16.16, Q8.16, and Q8.8 preserve functional autoregressive behavior, while Q4.4 falls below the minimum precision required for stable inference. Among the evaluated configurations, Q8.8 provides the most favorable trade-off between numerical fidelity and hardware efficiency.

\subsection{Micro-Phase Latency Profiling and Bottleneck Analysis}
\begin{figure}
    \vspace*{-1.5cm}
    \centering
    \includegraphics[width=1\linewidth]{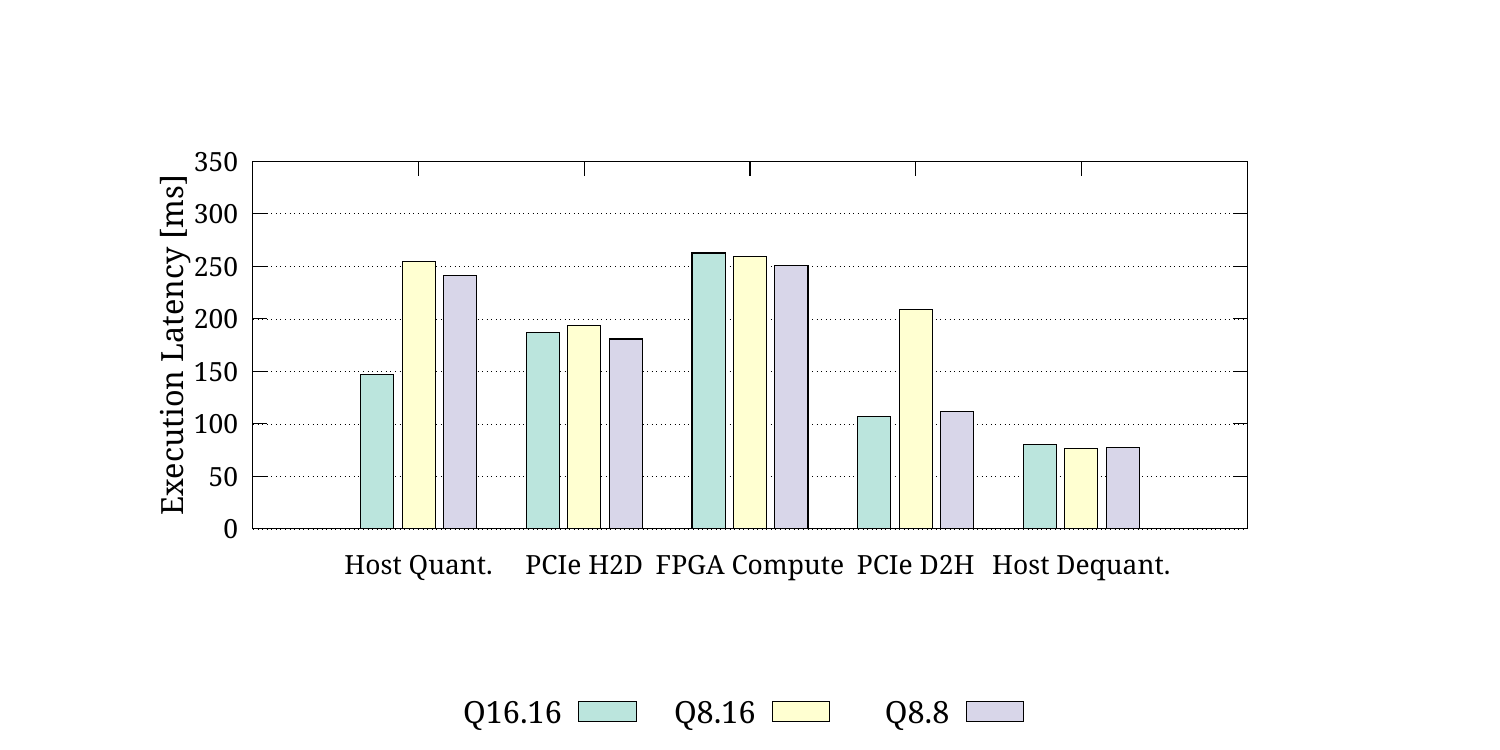}
    \caption{Total accumulated latency for 2,400 accelerator invocations, decomposed by pipeline phase ($T_{quant}$, $T_{H2D}$, $T_{kernel}$, $T_{D2H}$, $T_{dequant}$).}
    \label{fig:latency}
\end{figure}

Fig.~\ref{fig:latency} summarizes the latency measurements collected from 2,400 accelerator invocations corresponding to 24 Transformer layers evaluated over a 100-token autoregressive generation task. The FPGA kernel execution time remains nearly constant across all evaluated precisions, ranging from 251.16 ms to 262.67 ms for the complete workload, corresponding to approximately 104--109 $\mu$s per invocation. This behavior reflects the static pipelined architecture of the accelerator, whose execution schedule is independent of numerical precision. The Q16.16 configuration requires 147.01 ms, while Q8.16 and Q8.8 require 254.33 ms and 241.22 ms, respectively. The additional latency originates from clipping, overflow checking, and data conversion operations required for narrower fixed-point representations. The most significant observation is that system performance is dominated by communication overhead rather than computation. PCIe transfers account for approximately 34--41\% of the total execution time, while the combined cost of communication, quantization, and dequantization reaches 66.5\%, 73.9\%, and 70.9\% of the total runtime for Q16.16, Q8.16, and Q8.8, respectively. These results indicate that the proposed accelerator reduces non-linear function evaluation to the microsecond scale, but the overall system speed-up is constrained by host-device communication and runtime management overheads. Consequently, future performance improvements should focus on reducing data movement and increasing on-device processing. The overhead associated with loading new interpolation tables was not independently characterized in this work, as it includes software runtime, host-device communication, and XRT driver overhead in addition to the coefficient transfer itself. Since coefficient loading occurs only when switching between non-linear functions and the same tables are reused across multiple kernel invocations, this overhead is expected to be amortized during inference.

\section{Conclusion}
This work presented a general-purpose FPGA accelerator for non-linear function evaluation based on non-uniform piecewise-linear approximation. Unlike function-specific accelerators, the proposed architecture supports multiple non-linear functions via runtime-configurable coefficients while preserving a common hardware datapath, without requiring resynthesis or hardware modifications. Experimental results demonstrated adequate approximation of exponential, GELU, ReLU, Sigmoid, Swish, and tanh functions across multiple fixed-point precision formats. Evaluation using Transformer inference showed that the Swish accelerator maintains stable numerical behavior for the Q16.16, Q8.16, and Q8.8 configurations. On the WikiText-2 benchmark, these formats produced only minor degradation in loss and perplexity while preserving coherent autoregressive generation. In contrast, the Q4.4 configuration exceeded the numerical stability limit and resulted in significant performance degradation, identifying Q8.8 as the most attractive trade-off between hardware efficiency and inference quality. System-level profiling revealed that PCIe transfers and host-side data conversion dominate overall execution time, indicating that communication overhead remains the primary bottleneck in heterogeneous FPGA inference systems. Overall, the results demonstrate that non-uniform PWL approximation provides an effective balance between accuracy, flexibility, and hardware efficiency for FPGA-based neural network acceleration. Future work will focus on reducing host-device communication overhead, improving scalability to larger language models, exploring adaptive approximation strategies, extending the framework to additional Transformer operators, and performing comprehensive comparisons against optimized CPU and GPU implementations.

\section*{Acknowledgments}

This work was supported by the Costa Rica Institute of Technology under the research project 1360058 (Generación Automática de Hardware para Aplicaciones de Aprendizaje Automático basadas en FPGA). The results of this work were partially supported by AMD through the Heterogeneous Accelerated Compute Clusters (HACC) program. In addition, AI-based tools were used exclusively for grammar correction and translation during the preparation of this manuscript.

%
%
%
\bibliographystyle{splncs04}
\bibliography{mybibliography}

@article{haenlein2019brief,
  title={A Brief History of Artificial Intelligence: On the Past, Present, and Future of Artificial Intelligence},
  author={Haenlein, Michael and Kaplan, Andreas},
  journal={California Management Review},
  volume={61},
  number={4},
  pages={5--14},
  year={2019}
}

@inproceedings{vaswani2017attention,
  author    = {Ashish Vaswani and Noam Shazeer and Niki Parmar and Jakob Uszkoreit and Llion Jones and Aidan N. Gomez and {\L}ukasz Kaiser and Illia Polosukhin},
  title     = {Attention is All You Need},
  booktitle = {Advances in Neural Information Processing Systems (NeurIPS)},
  year      = {2017},
  pages     = {5998--6008}
}

@inproceedings{wei2020softmax,
  title={Design Space Exploration for Softmax Implementations},
  author={Wei, Z. and Arora, A. and Patel, P. and John, L.},
  booktitle={2020 IEEE 31st International Conference on Application-Specific Systems, Architectures and Processors (ASAP)},
  pages={45--52},
  year={2020},
  organization={IEEE},
  doi={10.1109/ASAP49362.2020.00017},
  address={Manchester, UK},
  month={Jul}
}

@article{kachris2025survey,
  title={A Survey on Hardware Accelerators for Large Language Models},
  author={Kachris, Christos},
  journal={Applied Sciences},
  volume={15},
  number={2},
  pages={586},
  year={2025},
  publisher={MDPI},
  doi={10.3390/app15020586},
}

@manual{amd_xilinx_xrt,
  title        = {Xilinx Runtime (XRT) Architecture and Python Bindings (PyXRT) Documentation},
  author       = {{AMD Xilinx}},
  year         = {2021},
  url          = {https://xilinx.github.io/XRT/2021.2/html/pyxrt.html},
}

@book{kastner2018parallel,
  title={Parallel Programming for FPGAs},
  author={Kastner, Ryan and Matai, Janarbek and Neuendorffer, Stephen},
  year={2018},
  publisher={Morgan Kaufmann}
}

@inproceedings{gupta2015deep,
  title     = {Deep Learning with Limited Numerical Precision},
  author    = {Gupta, Suyog and others},
  booktitle = {Proceedings of the 32nd International Conference on Machine Learning},
  series    = {Proceedings of Machine Learning Research},
  year      = {2015},
  publisher = {PMLR},
}

@INPROCEEDINGS{leiva-softmax,
  author={{Leiva-Valverde, Anthony and others}},
  booktitle={2026 IEEE 17th Latin America Symposium on Circuits and System (LASCAS)}, 
  title={{A Quantitative Evaluation of Approximate Softmax Functions for Deep Neural Networks}}, 
  year={2026},
  volume={},
  number={},
  pages={1-5},
  doi={10.1109/LASCAS67804.2026.11457137}}

@book{muller2016elementary,
  title={Elementary Functions: Algorithms and Implementation},
  author={Muller, Jean-Michel},
  edition={3rd},
  year={2016},
  publisher={Birkh{\"a}user},
  doi={10.1007/978-1-4899-7983-4}
}

@article{lachowicz2008fpga,
  title={Fast evaluation of nonlinear functions using {FPGA}s},
  author={Lachowicz, S. and Pfleiderer, H.-J.},
  journal={Advances in Radio Science},
  volume={6},
  pages={233--237},
  year={2008},
  doi={10.5194/ars-6-233-2008}
}

@INPROCEEDINGS{llm_fpga,
  author={Prieto-Sibaja, Luis D. and others},
  booktitle={2024 IEEE 42nd Central America and Panama Convention (CONCAPAN XLII)}, 
  title={LLM Acceleration on FPGAs: A Comparative Study of Layer and Spatial Accelerators}, 
  year={2024},
  volume={},
  number={},
  pages={1-6},
  doi={10.1109/CONCAPAN63470.2024.10933896}}

@article{carrillo2025cordic_softmax,
  title={{Evaluating CORDIC-Based and Approximate Softmax Accelerators for Deep Learning Inference on FPGAs}},
  author={{Carrillo-Arroyo, Andrés and others}},
  journal={Accepted in JOCICI 2026},
  year={2026}
}

@manual{amd_dataflow_canonical,
  title        = {Vitis HLS User Guide: Dataflow Canonical Rules},
  author       = {{Advanced Micro Devices (AMD)}},
  organization = {AMD},
  year         = {2024},
  note         = {UG1399 and UG1448 documentation},
  url          = {https://docs.amd.com/r/en-US/ug1448-hls-guidance/Dataflow-Canonical-Rules-2}
}

@manual{xilinx_hls,
  title={{Vitis High-Level Synthesis User Guide (UG1399)}},
  author={{AMD Xilinx}},
  year={2023},
  version={v2023.2},
  organization={Advanced Micro Devices, Inc.}
}

@article{chen2025fpga,
  title={Understanding the Potential of {FPGA}-Based Spatial Acceleration for Large Language Model Inference},
  author={Chen, Hongwu and others},
  journal={ACM Transactions on Reconfigurable Technology and Systems},
  volume={18},
  number={1},
  year={2026},
  publisher={ACM New York, NY, USA},
  doi={10.1145/3656177}
}

@book{burden2015numerical,
  author    = {Burden, Richard L. and Faires, J. Douglas and Burden, Annette M.},
  title     = {Numerical Analysis},
  edition   = {10th},
  publisher = {Cengage Learning},
  address   = {Boston, MA, USA},
  year      = {2015},
  isbn      = {978-1-305-25366-7}
}

@article{mahbub2026hardware_journal,
  title     = {Hardware Accelerators for Softmax in Large Language Models: A Survey},
  author    = {Mahbub, Raaida Noor and others},
  journal   = {The Journal of Supercomputing},
  year      = {2026},
  doi       = {10.1007/s11227-026-08322-x},
  url       = {https://link.springer.com/article/10.1007/s11227-026-08322-x}
}

\end{document}